\documentclass[aps, prd, amsmath, floats, floatfix, twocolumn, nofootinbib, showpacs]{revtex4-1}
\usepackage{graphicx}
\usepackage{color}
\usepackage{latexsym}

\newcommand{\be}{\begin{eqnarray}}
\newcommand{\ee}{\end{eqnarray}}

\begin{document}

\title{X-ray spectropolarimetric signature of a warped disk around a stellar-mass black hole}

\author{Yifan Cheng}

\author{Dan Liu}

\author{Sourabh Nampalliwar}

\author{Cosimo Bambi}
\email[Corresponding author: ]{bambi@fudan.edu.cn}

\affiliation{Center for Field Theory and Particle Physics and Department of Physics, Fudan University, 200433 Shanghai, China}

\date{\today}

\begin{abstract}
Black holes in X-ray binaries are often assumed to be rotating perpendicular to the plane of the accretion disk and parallel to the orbital plane of the binary. While the Bardeen-Petterson effect forces the inner part of the accretion disk to be aligned with the equatorial plane of a spinning black hole, the disk may be warped such that the inclination angle of the outer part is different from that of the inner part. In this paper, we identify a possible observational signature of a warped accretion disk in the spectrum of the polarization degree of the continuum. Such a signature would provide direct evidence for the presence of a warped disk and, potentially, even a measure of the warp radius, which, in turn, could be used to infer the viscosity parameter of the disk.
\end{abstract}


\maketitle


\section{Introduction \label{s-1}}

Accretion disks around black holes (BH hereafter) are an important tool for studying the physics and the astrophysics in the vicinity of BHs~\cite{snz}. The inner edge of an accretion disk is typically very close to the central BH and the radiation released by the accreting gas gets affected by the strong gravity of the BH. Thus, radiation emanating from the accretion disk and observed far away from the BH encodes information about the spacetime geometry near the BH. A study of the properties of this radiation is useful, among other things, to measure the spin of BHs~\cite{spin} or even to test general relativity~\cite{kerr}. In this paper, we analyze the effect of the BH on the polarization of radiation as it emerges from the disk and arrives at a far away observer. We are interested in BHs found in X-ray binary systems. Observational data with the polarization information of such systems are not available today, but such observations hold promise in the near future, for instance with the China-Europe mission eXTP~\cite{xtp}, the European project XIPE~\cite{xipe}, or the two NASA missions recently approved for phase-A study: IXPE~\cite{ixpe} and PRAXYS~\cite{praxys}.

 It is usually assumed that the BH spin is aligned with the orbital angular momentum vector of the binary system, but this is not obvious. Such an assumption plays a crucial role, for instance, in the measurements of the BH spin from the continuum~\cite{cfm}: the continuum-fitting method cannot determine the inclination angle of the inner part of the disk because there is a degeneracy between this angle and the spin parameter. The disk's inclination angle is usually inferred from optical observations of the ellipsoidal modulation in the light curve of the companion star~\cite{orosz}. However, the latter measures the inclination angle of the orbital plane of the system and of the outer part of the accretion disk.

The inner part of the accretion disk is expected to be on the plane perpendicular to the BH spin as a result of the Bardeen-Petterson effect~\cite{bp}. The mechanism should work for thin disks, because it requires that $\alpha > h/r$, where $\alpha \sim 0.01-0.1$ is the viscosity parameter and $h/r$ is the disk semi-thickness. The combination of the Lense-Thirring precession with the disk viscosity eventually drags the innermost part of the disk into alignment with the BH spin. Because of the short range of the Lense-Thirring effect, the outer part tends to remain in its original configuration. (For more details, see e.g.~\cite{bp,ppp}.) The ``Bardeen-Petterson configuration'' refers to a system in which the inner part of the disk is flat and perpendicular to the BH spin, while the outer part is also flat but in the plane perpendicular to the angular momentum vector of the binary. Here we are interested in this configuration. For instance, if the inner part of the disk is a hot geometrically thick accretion flow, the picture is different and the inner disk precesses as a solid body~\cite{ingram}.

The alignment timescale of thin disks has been estimated to be in the range $10^6$-$10^8$~yrs, and therefore the disk should be already adjusted in the BH equatorial plane for not too young systems~\cite{jack}. However, the actual timescale depends on some unknown parameters like the viscosity $\alpha$~\cite{r_w} and it should be note that at least some numerical simulations do not find the adjustment of the disk~\cite{fragile}. Since the disk is created from the material of the stellar companion, the outer part of the disk is necessarily in the orbital plane of the binary system. The question is thus if the planes of the inner disk and of the outer disk coincide. If the two portions are not on the same plane, the disk is warped.

Theoretical arguments suggest that both warped and unwarped disks are possible. This mainly depends on the origin of the binary system. If the system formed through multi-body interactions, such as binary capture or replacement, the orientations of the BH spin and of the orbital angular momentum of the binary are completely uncorrelated and it is quite unlikely to have them aligned. If the BH formed from the supernova explosion of a heavy star in a binary, its spin would be more likely aligned with the orbital angular momentum vector of the system in the case of a symmetric explosion without strong shocks and kicks, while a misalignment would be created otherwise~\cite{fragos2010}. Current attempts to test the possibility of the presence of a warped disk rely on the comparison between the estimate of the inclination angle of radio jets in microquasar, under the assumption that the jet is aligned with the BH spin, and optical measurements of the light curve of the binary, which are sensitive to the inclination angle of the outer part of the disk. At present, there may be evidence for both aligned and misaligned disks~\cite{jack, mac02}. However, this approach cannot measure the warp radius and it is not guaranteed that the inner part of the disk is in the plane perpendicular to the BH spin, as predicted by the Bardeen-Petterson effect.

The aim of this paper is to explore the possibility of identifying a warped disk with X-ray spectropolarimetric measurements of the thermal spectrum of the accretion disk, the so-called continuum. We will focus our attention on stellar-mass BHs in X-ray binary systems, as we are interested in X-ray spectropolarimetric observations of the thermal spectrum of thin disks. Such spectropolametric measurements are limited in scope in the case of supermassive  BHs in galactic nuclei, due to extinction and dust absorption affecting the thermal spectrum of the accretion disk.

The thermal spectrum of a disk is initially unpolarized, but it becomes weakly polarized due to Thomson scattering of X-ray photons off free electrons in the dense atmosphere above the disk. The observed polarization degree and polarization angle are strongly affected by the propagation of the photons in the strong gravitational field of the BH and therefore it is strictly necessary to perform these calculations in the framework of general relativity~\cite{stark,lixin,schnittman}. 
Here we employ a simple model to describe a warped disk and we find that it is possible to identify a specific signature in the spectrum of the polarization degree. The sole spectrum of the polarization degree can identify the presence of a warped disk and possibly determine its warp radius $R_*$, namely the transition radius between the inner and the outer parts. The possibility of this kind of measurement eventually depends on the properties of the accretion disk, namely the viewing angles of the inner and the outer disks and the value of the warp radius, and the sensitivity and the energy band of the polarimetric detector. For the first generation of X-ray polarimetric missions, which will probably work in the 1-10~keV range, the detection of a warped disk will only be possible in the case of small warp radii, $R_* < 100$~$R_g$, where $R_g = G_N M / c^2$ is the BH gravitational radius.

The contents of this paper is as follows. In Section~\ref{s-2}, we briefly review the calculations of the  polarization properties of a geometrically thin and optically thick accretion disk around a stellar-mass BH. In Section~\ref{s-3}, we consider a warped disk, in which the inclination angles of the inner and the outer parts are different, and we find a possible observational signature in the spectrum of the polarization degree. Summary and conclusions are reported in Section~\ref{s-4}.

\section{Polarization of the continuum \label{s-2}}

Geometrically thin and optically thick accretion disks around  BHs are commonly described with the Novikov-Thorne model~\cite{nt-model}. In this framework, the disk is supposed to be in the plane perpendicular to the BH spin, the particles of the gas follow nearly geodesic circular orbits, and the inner edge of the disk is at the innermost stable circular orbit, or ISCO. From the conservation laws of rest-mass, energy, and angular momentum, it is possible to determine the time-averaged radial structure of the disk and the time-averaged energy flux from the disk surface $\mathcal{F}(r)$~\cite{nt-model}. Since the disk is in thermal equilibrium, the emission is blackbody-like and we can define an effective temperature $T_{\rm eff}(r) = (\mathcal{F}/\sigma_{\rm SB})^{1/4}$, where $\sigma_{\rm SB}$ is the Stefan-Boltzmann constant. Electron scattering in the atmosphere can be taken into account by introducing the color temperature $T_{\rm col} = f T_{\rm eff}$, where $f$ is the hardening factor (e.g., for stellar mass  BHs accreting around 10\% the Eddington rate $f = 1.6$). The local specific intensity of the radiation emitted by the disk is
\be
I_{\rm e} (\nu_{\rm e}) = \frac{2 h \nu_{\rm e}^3}{c^2} \frac{\Upsilon}{f^4} \left[\exp\left(\frac{h \nu_{\rm e}}{k_{\rm B} T_{\rm col}}\right) - 1\right]^{-1} \, ,
\ee
where $\nu_{\rm e}$ is the photon frequency (the subindex ``e'' refers to quantities measured in the rest frame of the emitter), $h$ is the Planck constant, $c$ is the speed of light, $k_{\rm B}$ is the Boltzmann constant, and $\Upsilon$ is a function of the angle between the photon propagation direction and the normal to the disk surface (for the calculations, see e.g.,~\cite{code-c}).

The thermal radiation of the disk is initially unpolarized, but it becomes weakly polarized due to Thomson scattering of X-ray photons off free electrons in the atmosphere just above the disk~\cite{chandra}. The polarization degree $\delta$ only depends on the angle between the direction of propagation of the photon and the normal to the surface of the disk, ranging from 0\%, for photons with direction perpendicular to the disk, to about 12\%, for photons in the plane of the disk. The angle of polarization $\psi$ is instead perpendicular to the direction of propagation of the photons and parallel to the disk plane. More details can be found in~\cite{chandra}.

The strong gravitational field in the vicinity of the central BH affects the parallel transport of the polarization vector along the photon path and therefore the polarization properties of the radiation reaching a distant detector must be calculated in the framework of general relativity. The first studies of the polarization of thin accretion disks around  BHs were reported in~\cite{stark}, and later extended in~\cite{lixin,schnittman}. In the present paper, the calculations are done with the code described in~\cite{code-dan}, which is an extension of that in~\cite{code-c}. We compute the photon trajectories backward in time from every point in the plane of the distant observer to the emission point in the accretion disk and we determine the polarization degree and angle at the observer point as described in the appendix in~\cite{code-dan}. We then integrate over the plane of the distant observer to get the spectrum of $\delta_{\rm obs}$ and $\psi_{\rm obs}$. In terms of the Stokes parameters $I$, $Q$, $U$, and $V$~\cite{chandra}, for each point on the image we have
\be
Q + i U = \delta I e^{2 i \psi} \, ,
\ee
where $V=0$ because the radiation is linearly polarized. The radiation field is decomposed into a completely polarized component $I^p = \delta I$ and an unpolarized one $I^u = (1-\delta) I$. At the point of detection, we have
\be
\vspace{-0.8cm}
\langle Q_{\rm obs} \rangle + i \langle U_{\rm obs} \rangle &=&
\frac{1}{\Delta \Omega_{\rm obs}} \int \left( Q_{\rm obs} 
+ i U_{\rm obs} \right) d\Omega_{\rm obs}
\nonumber\\
&=&\frac{1}{\Delta \Omega_{\rm obs}} \int g^3 \delta_{\rm e} I_{\rm e}
e^{2 i \psi_{\rm obs}} d\Omega_{\rm obs} \, ,
\ee
where $\langle \cdot \rangle$ indicates the average over the image, $\Delta \Omega_{\rm obs}$ is the total solid angle subtended by the disk in the sky, and the redshift factor $g = E_{\rm obs}/E_{\rm e}$ follows from Liouville's theorem $I_{\rm obs}/E_{\rm obs}^3 = I_{\rm e}/E_{\rm e}^3$. In general, an initially completely polarized radiation is detected on the observer's plane as partially polarized, because different points of the image have photons with different $\psi_{\rm obs}$. The total intensity at the detection point is
\be
\langle I_{\rm obs} \rangle &=&
 \frac{1}{\Delta \Omega_{\rm obs}} \int
g^3 I_{\rm e} d\Omega_{\rm obs} \nonumber\\
&=& \langle I^u_{\rm obs} \rangle + \langle I^p_{\rm obs} \rangle \, .
\ee
The observed averaged polarization degree is~\cite{lixin}
\be\label{eq-delta-av}
\langle \delta_{\rm obs} \rangle &=& \, 
\frac{\sqrt{\langle Q_{\rm obs} \rangle^2 + \langle U_{\rm obs} 
\rangle^2}}{\langle I_{\rm obs} \rangle} \, , 
\ee
and the observed averaged polarization angle is determined from the following two relations~\cite{lixin}
\be
\sin \left( 2 \langle \psi_{\rm obs} \rangle \right) &=& \frac{\langle U_{\rm obs} 
\rangle}{\sqrt{\langle Q_{\rm obs} \rangle^2 + \langle U_{\rm obs} \rangle^2}} \, , \\
\cos \left( 2 \langle \psi_{\rm obs} \rangle \right) &=& \frac{\langle Q_{\rm obs} 
\rangle}{\sqrt{\langle Q_{\rm obs} \rangle^2 + \langle U_{\rm obs} \rangle^2}} \, .
\ee

In the present analysis, we ignore two features: ``returning'' radiation from the disk and photon absorption due to the disk atmosphere. To describe returning radiation, we first define ``direct'' radiation as that radiation which, once emitted from the accretion disk, arrives at the observer plane without re-absoprtion. Then, returning radiation is that radiation which suffers so much gravitational bending that it ``returns'' to the disk, gets re-absorbed and re-emitted, and then arrives at the observer plane. Most of the returning radiation comes from the disk region close to the BH, because of the presence of stronger gravitational field there. Since the primary contribution to returning radiation involves the radiation emitted in the inner part of the accretion disk, it affects the spectrum at high energies, e.g., $E > 3$~keV for a 10~$M_{\odot}$ BH accreting at about 10\% the Eddington rate (see the figures in Refs.~\cite{lixin,schnittman}).

The photon absorption due to the disk atmosphere reduces the polarization degree and it is important at large radii, equivalently, in the low energy ($E < 0.1$~keV) part of the spectrum (see again the figures in Refs.~\cite{lixin,schnittman}). As we will show in the next section, the observational signature of the presence of a warped disk should show up in the range 0.1-3~keV and therefore our results and conclusions are not expected to be affected by the simplifications in our model. Photon absorption by the interstellar medium between the source and the observer does not affect our calculations, because it is independent of the emission point on the disk and it can be modeled by multiplying the intensity by a function of energy, and therefore this correction drops out in Eq.~(\ref{eq-delta-av}).

Although we have ignored returning radiation, we must include radiation emitted by the inner disk and scattered by the outer disk (which would be present even in flat spacetime due to the different inclination angles of the two disks). We call this ``reflected'' radiation. The impact of the radiation in the opposite direction, i.e., emitted from the outer disk and scattered by the inner disk, can be neglected because the temperature, and therefore the photon flux, of the outer disk is lower than that of the inner disk.

\begin{figure*}
\begin{center}
\includegraphics[type=pdf,ext=.pdf,read=.pdf,width=8.9cm]{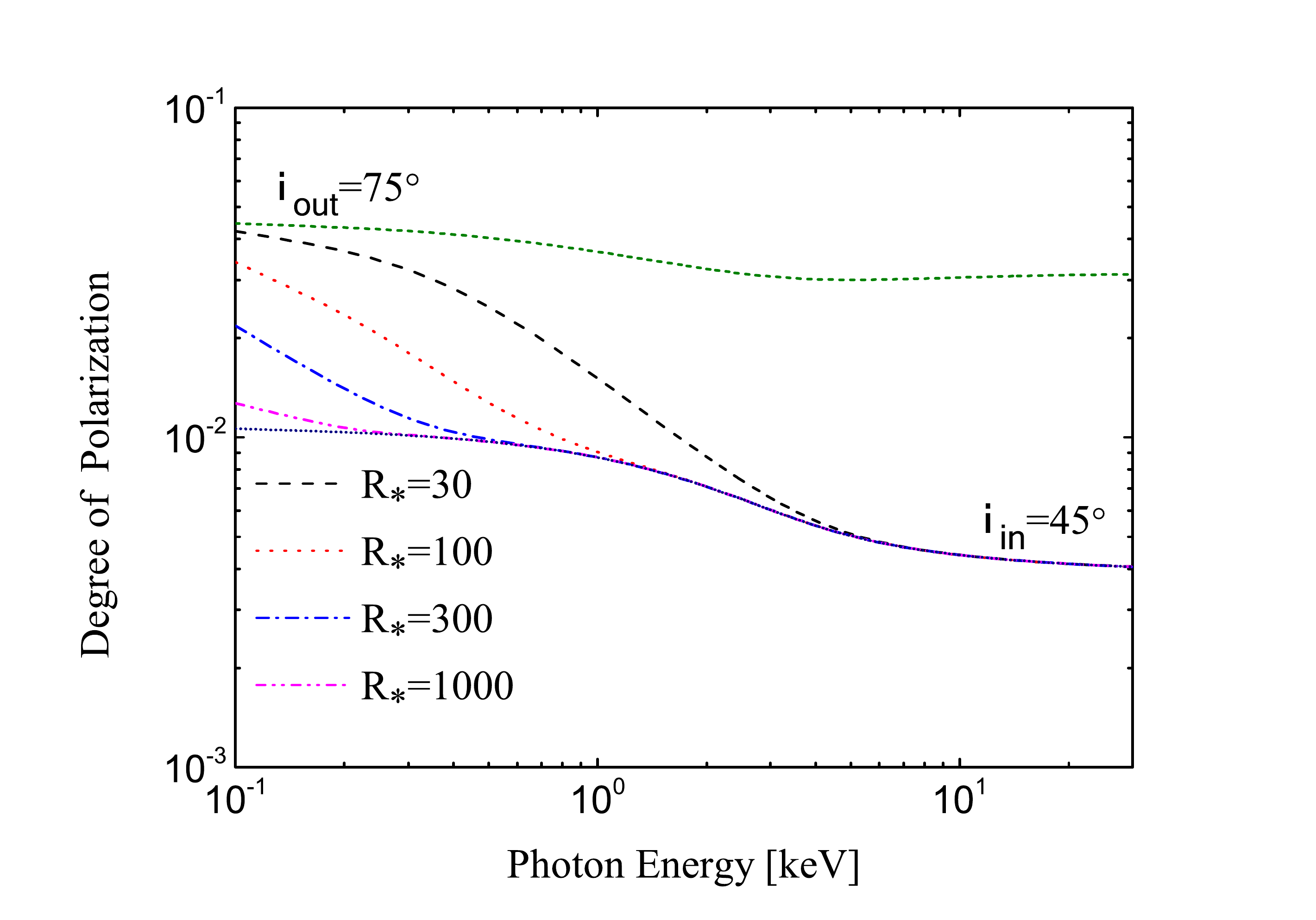}
\includegraphics[type=pdf,ext=.pdf,read=.pdf,width=8.9cm]{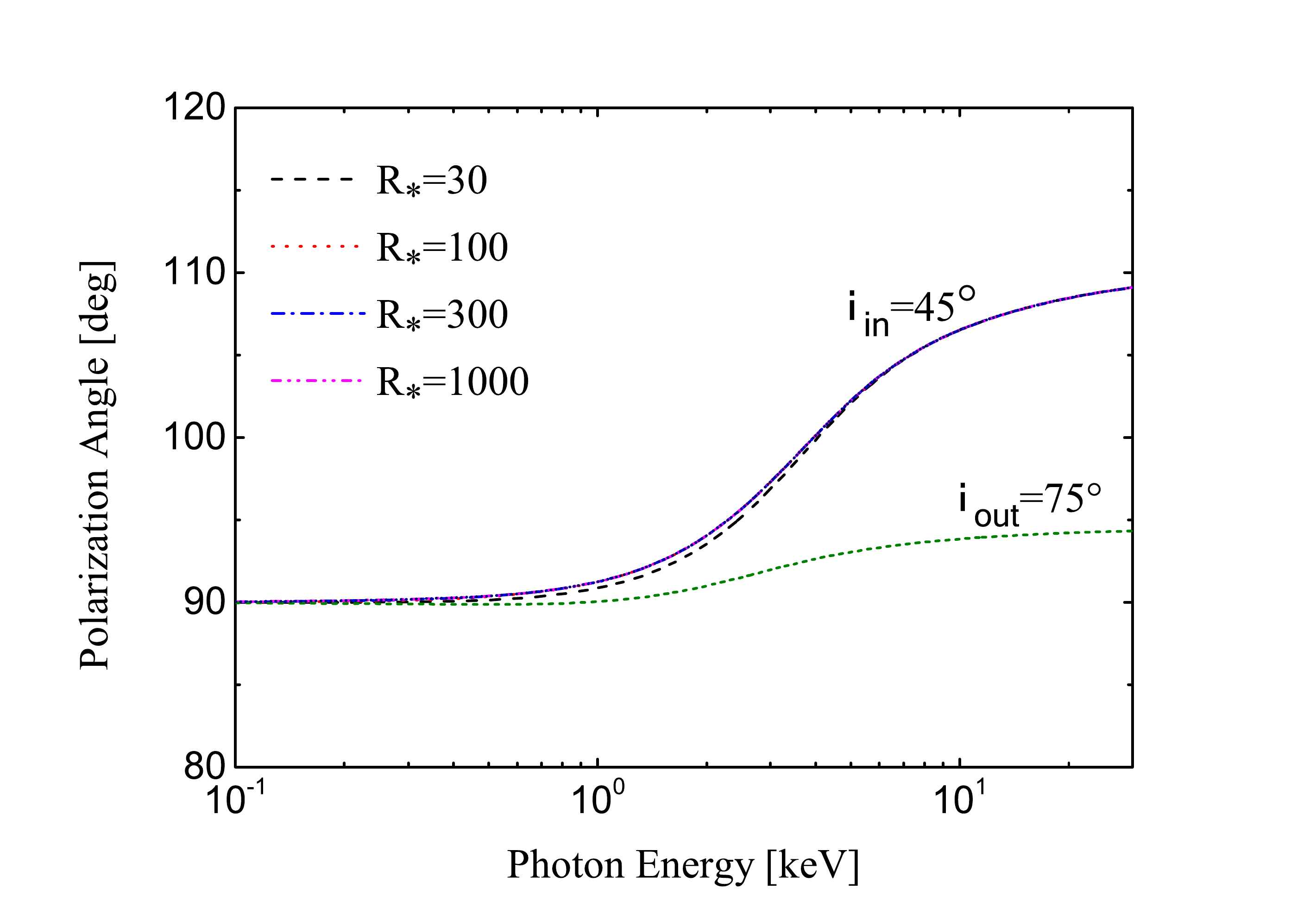}
\end{center}
\caption{Polarization degree (left panel) and polarization angle (right panel) as a function of the photon energy for the thermal spectrum of an accretion disk around a Schwarzschild BH without including reflected radiation. Here the inner disk has an inclination angle $i_{\rm in} =45^\circ$ with respect to the line of sight of the observer, while the outer part has $i_{\rm out} =75^\circ$. The warp radius is $R_*/R_g = 30$, 100, 300, 1000 and we show also the cases of unwarped disks with $i_{\rm in} = i_{\rm out} =45^\circ$ (blue dotted line) and $i_{\rm in} = i_{\rm out}=75^\circ$ (green dashed line). The BH mass is taken as $10$~$M_{\odot}$ with a luminosity of 10\% the Eddington limit. See the text for more details.}
\label{fig1}
\vspace{0.3cm}
\begin{center}
\includegraphics[type=pdf,ext=.pdf,read=.pdf,width=8.9cm]{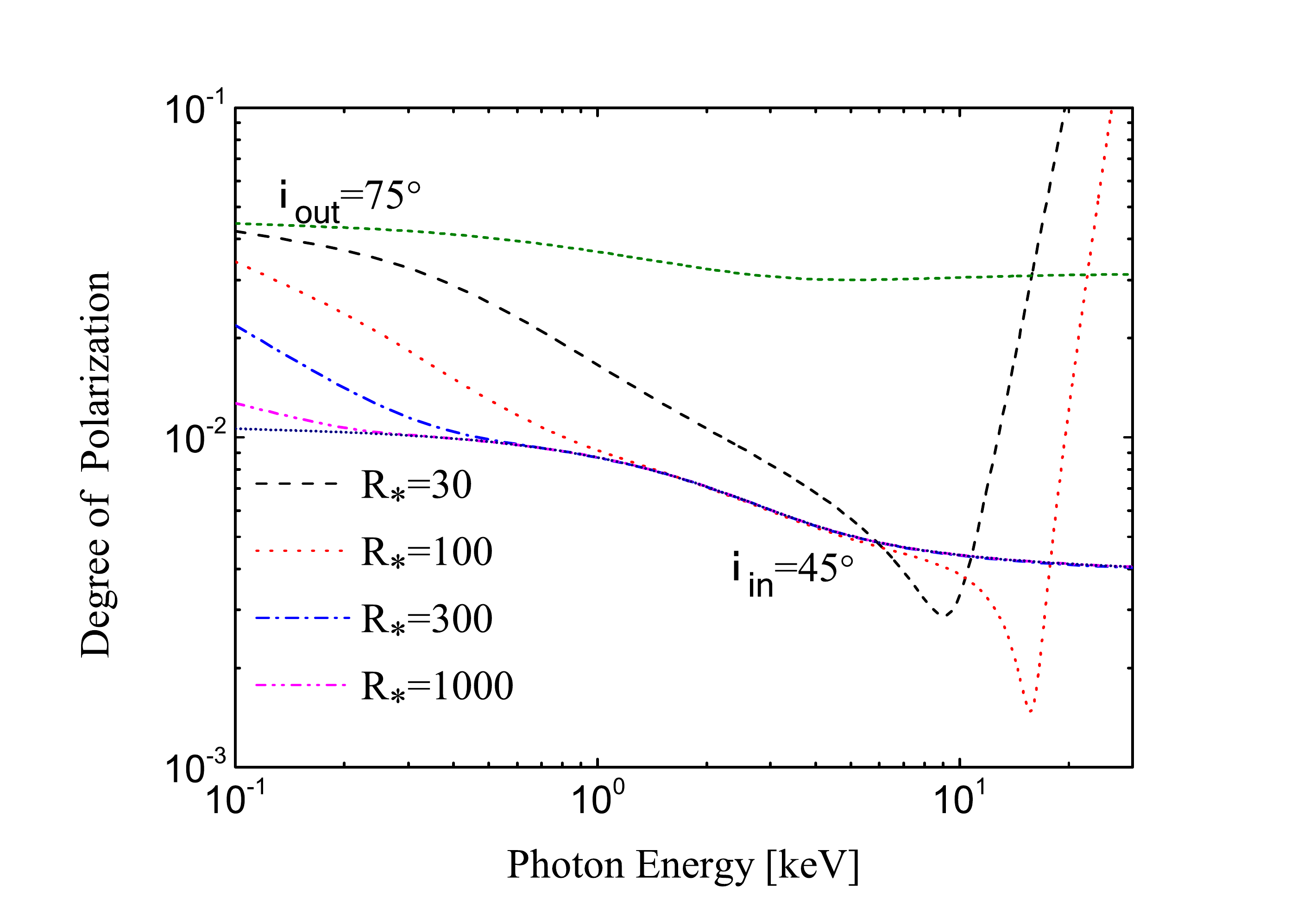}
\includegraphics[type=pdf,ext=.pdf,read=.pdf,width=8.9cm]{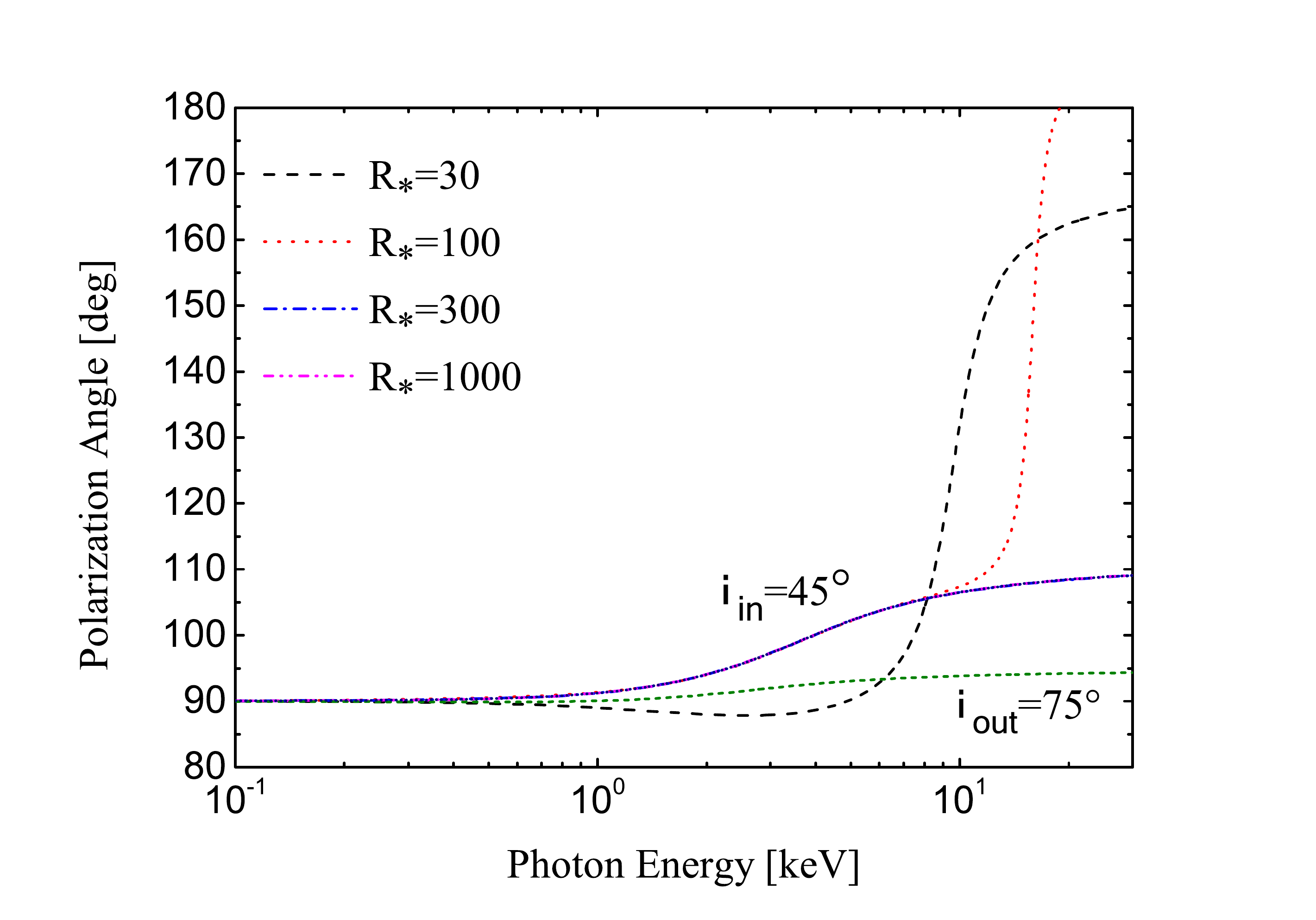}
\end{center}
\caption{As in Fig.~\ref{fig1}, but including reflected radiation.}
\label{fig1w}
\end{figure*}

\begin{figure*}
\begin{center}
\includegraphics[type=pdf,ext=.pdf,read=.pdf,width=8.9cm]{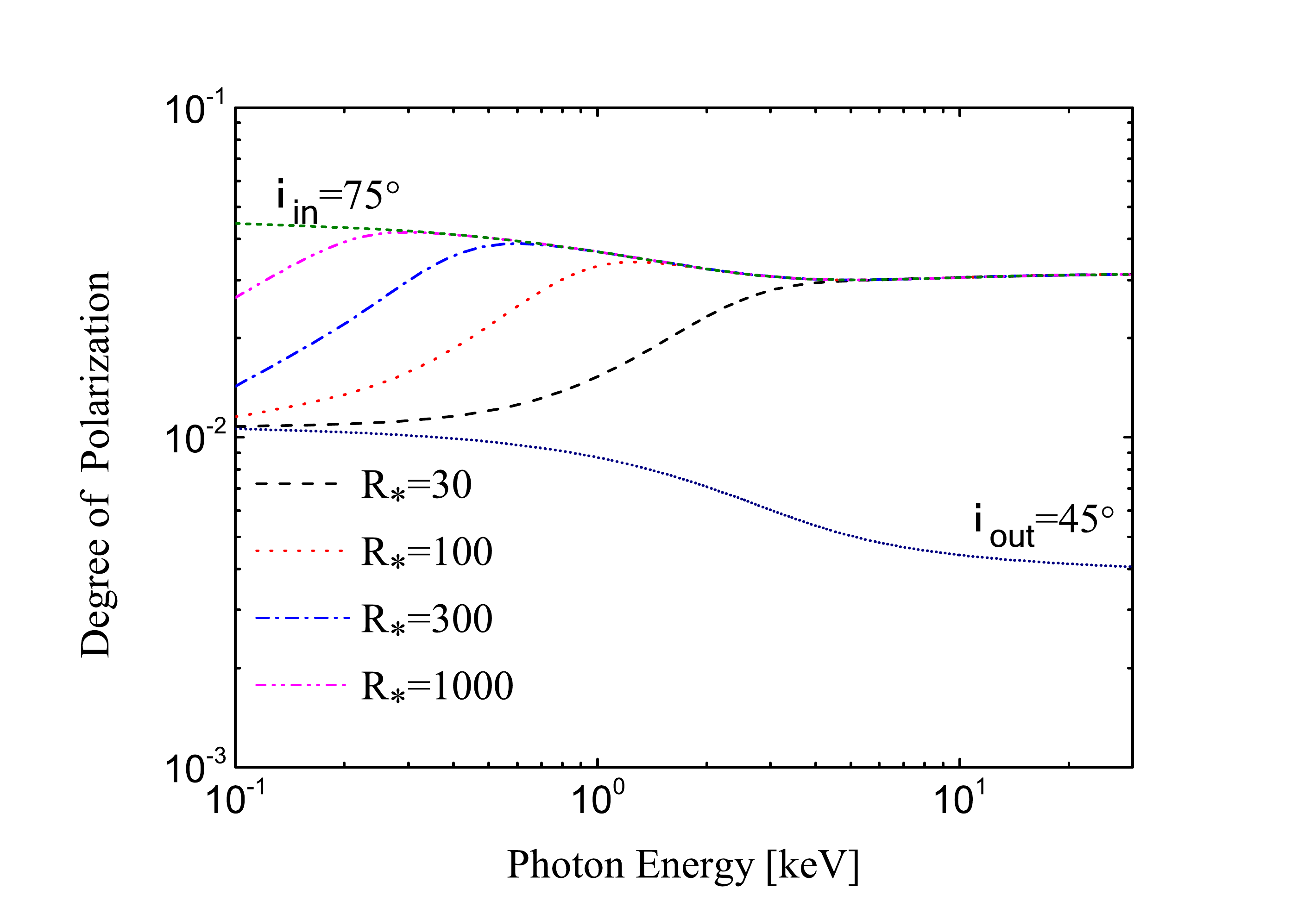}
\includegraphics[type=pdf,ext=.pdf,read=.pdf,width=8.9cm]{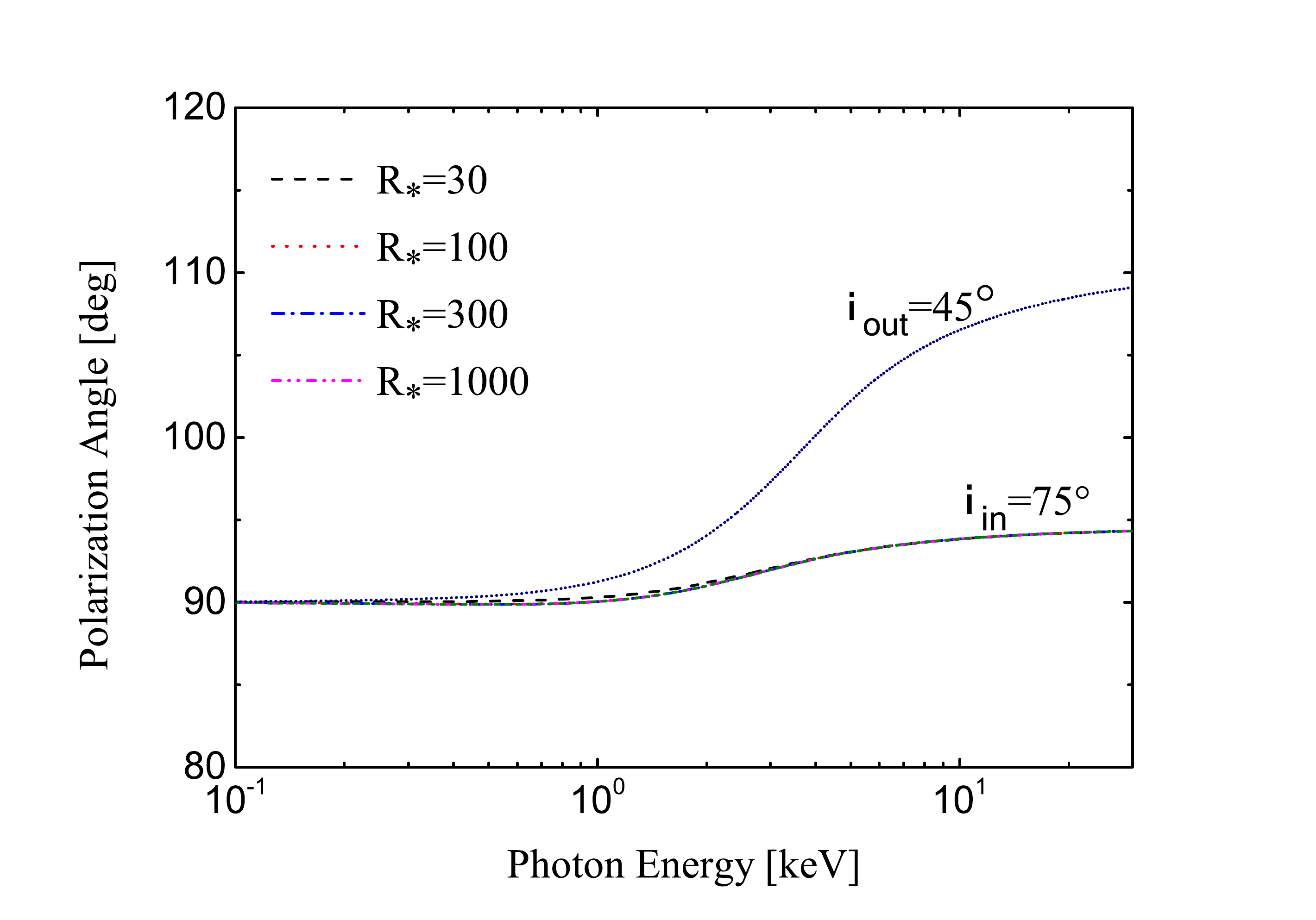}
\end{center}
\caption{As in Fig.~\ref{fig1}, but now the inner part of the disk has an inclination angle $i_{\rm in}=75^\circ$ with respect to the line of sight of the observer, while the outer part has $i_{\rm out}=45^\circ$. See the text for more details.}
\label{fig2}
\vspace{0.3cm}
\begin{center}
\includegraphics[type=pdf,ext=.pdf,read=.pdf,width=8.9cm]{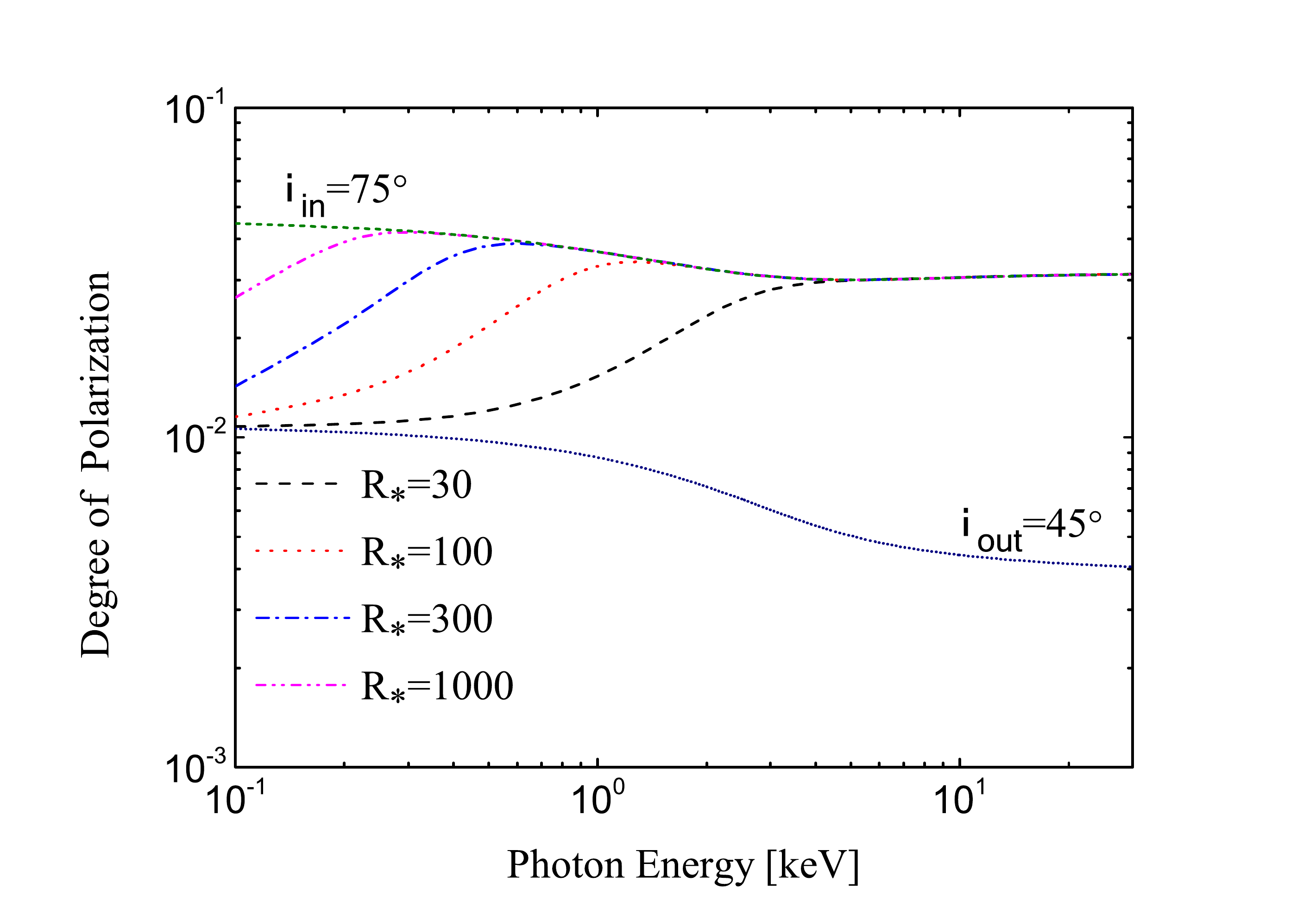}
\includegraphics[type=pdf,ext=.pdf,read=.pdf,width=8.9cm]{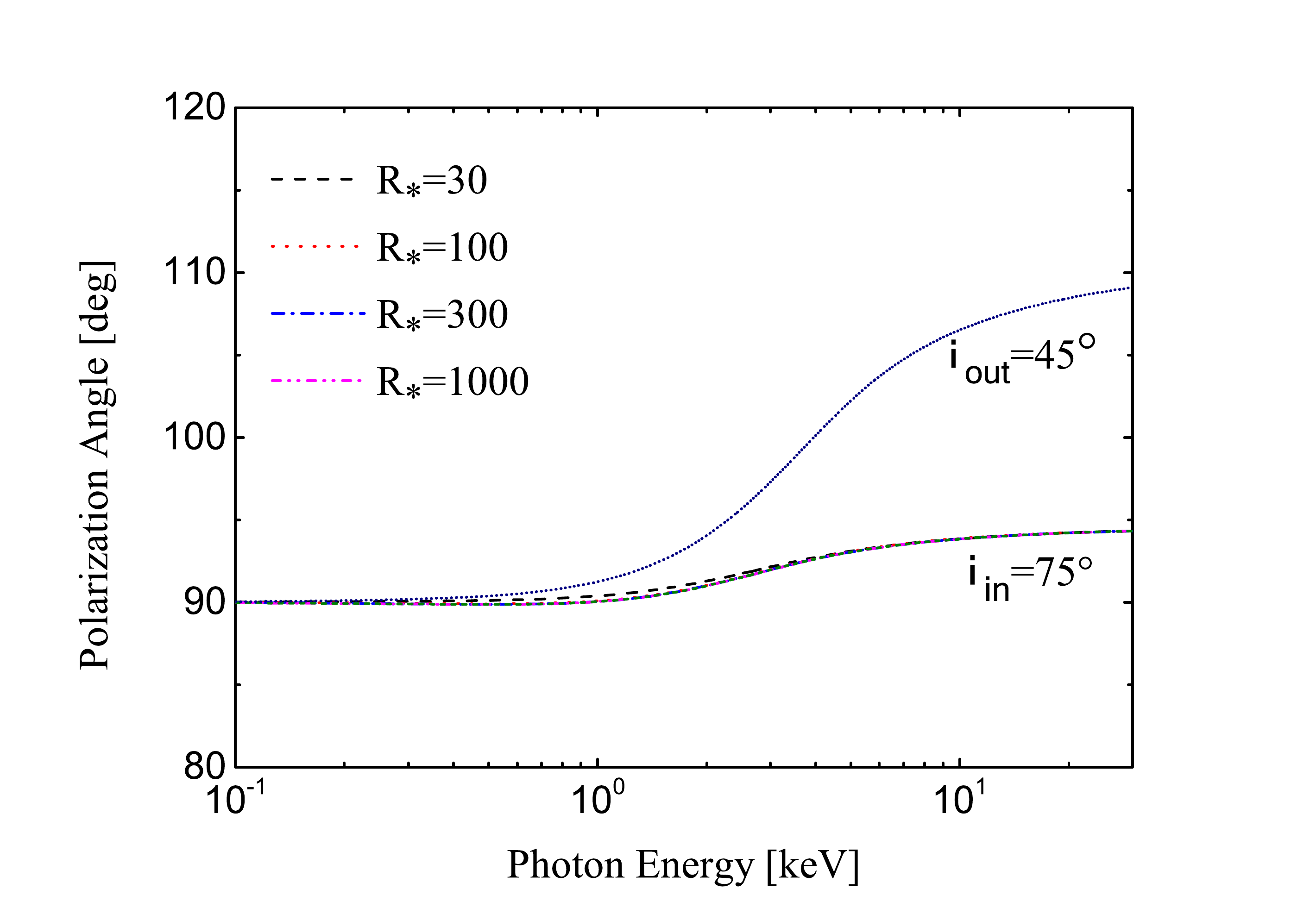}
\end{center}
\caption{As in Fig.~\ref{fig2}, but including the radiation emitted from the inner disk and scattered by the outer disk.}
\label{fig2w}
\end{figure*}

\begin{figure*}
\begin{center}
\includegraphics[type=pdf,ext=.pdf,read=.pdf,width=8.9cm]{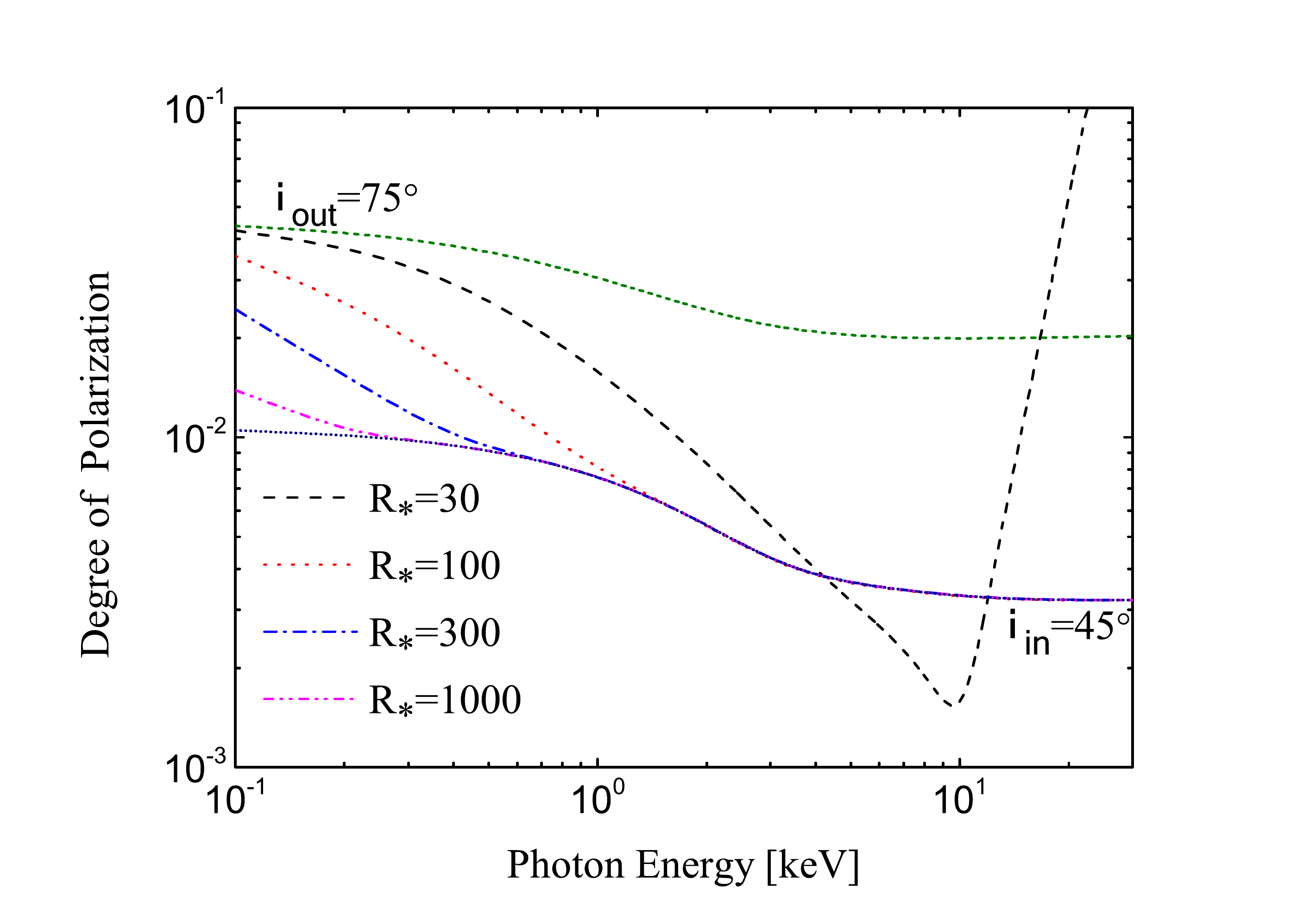}
\includegraphics[type=pdf,ext=.pdf,read=.pdf,width=8.9cm]{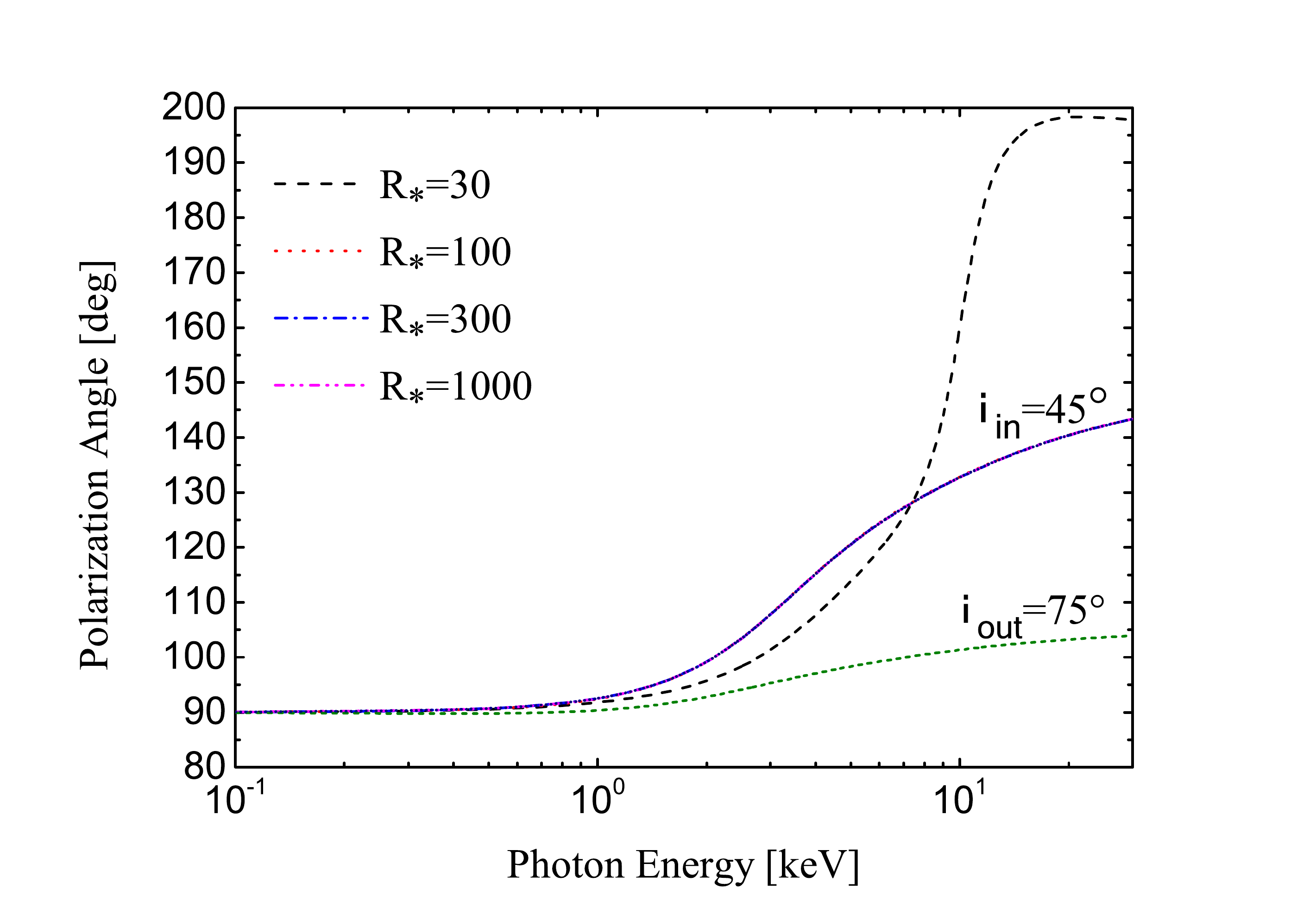}
\vspace{0cm}
\includegraphics[type=pdf,ext=.pdf,read=.pdf,width=8.9cm]{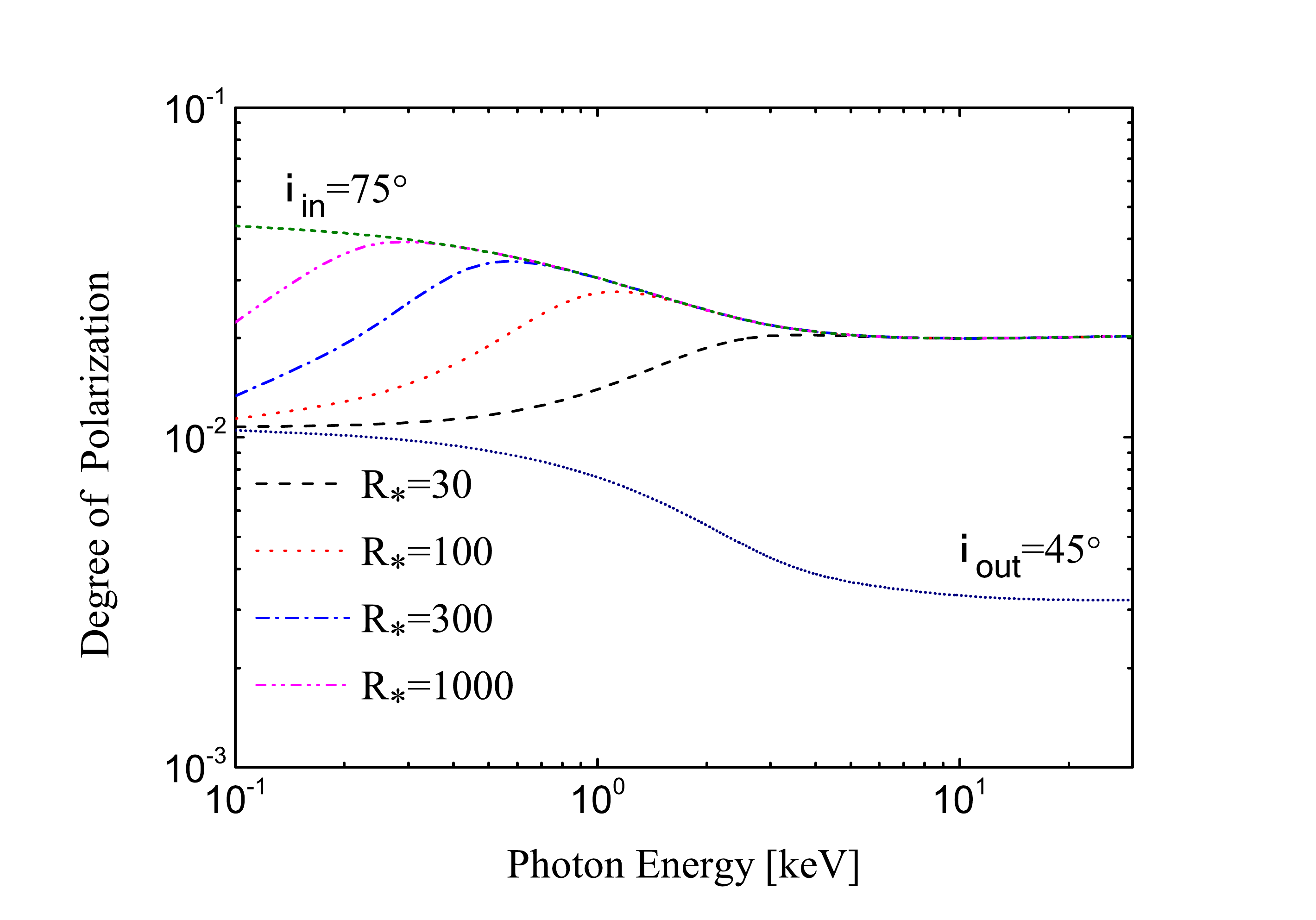}
\includegraphics[type=pdf,ext=.pdf,read=.pdf,width=8.9cm]{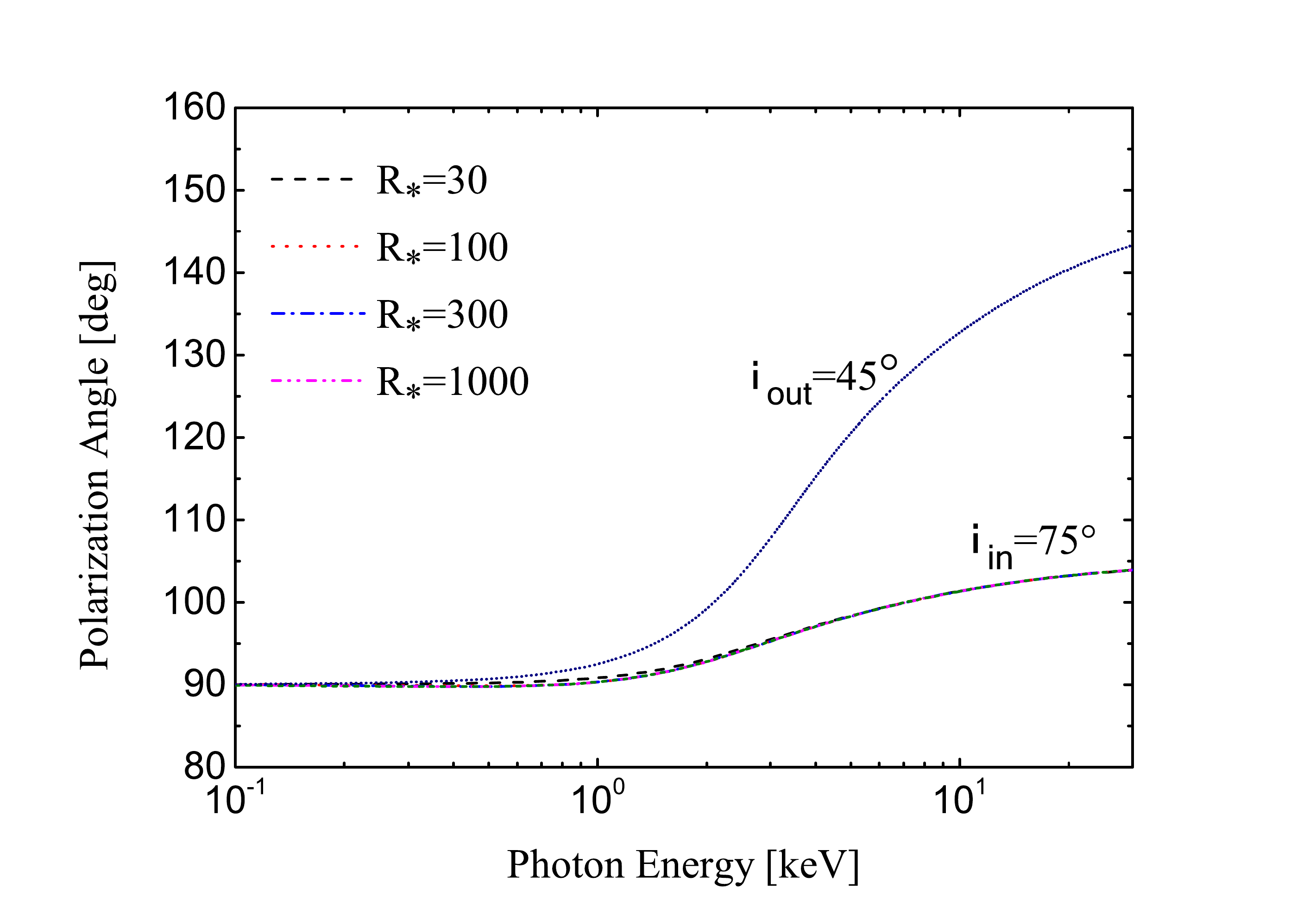}
\end{center}
\caption{Top panels: as in Fig.~\ref{fig1w} for the accretion disk around a Kerr BH with spin parameter $a_* = 0.9$. Bottom panels: as in Fig.~\ref{fig2w} for the accretion disk around a Kerr BH with spin parameter $a_* = 0.9$. See the text for more details.}
\label{fig3}
\vspace{0.3cm}
\begin{center}
\includegraphics[type=pdf,ext=.pdf,read=.pdf,width=8.9cm]{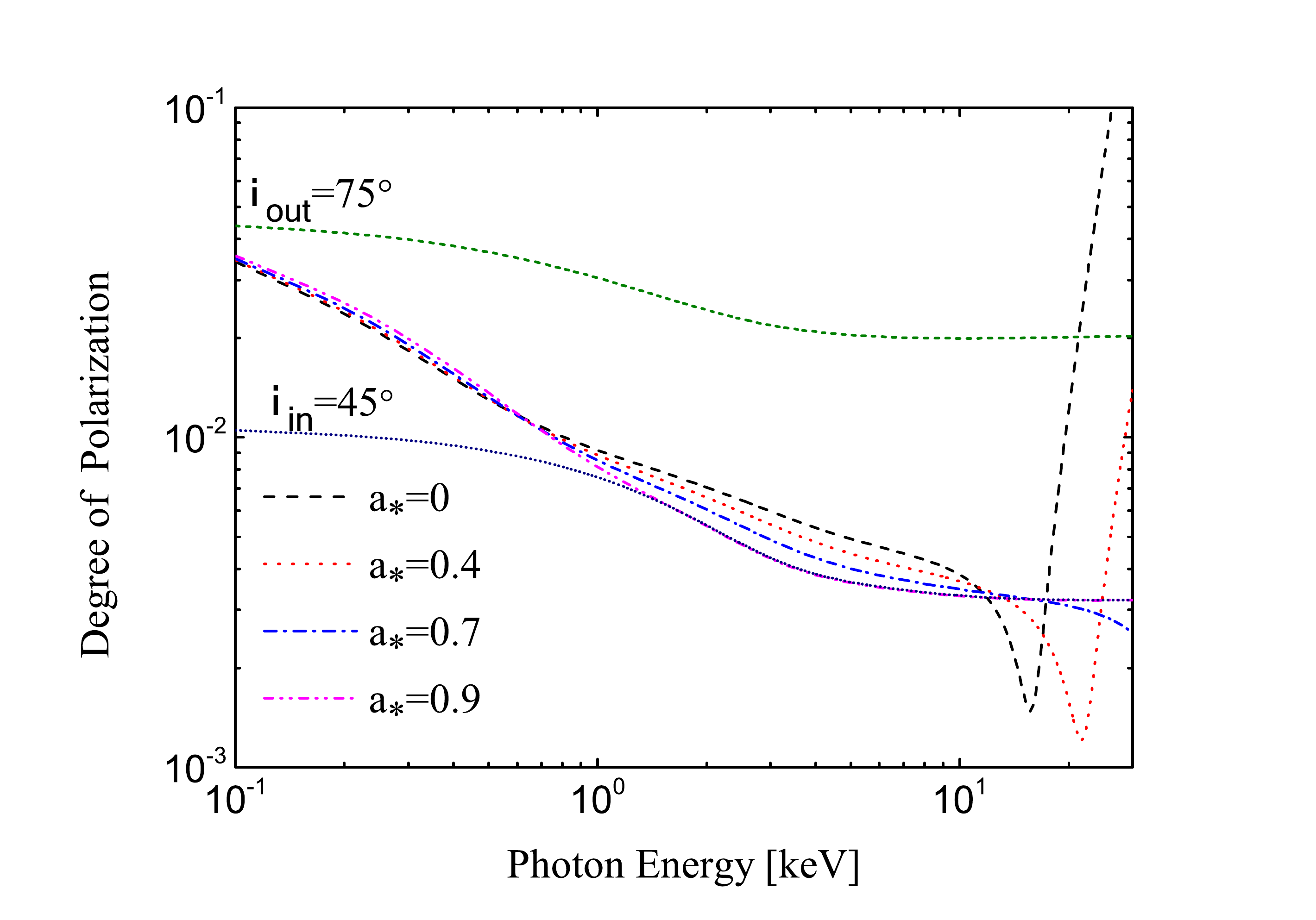}
\includegraphics[type=pdf,ext=.pdf,read=.pdf,width=8.9cm]{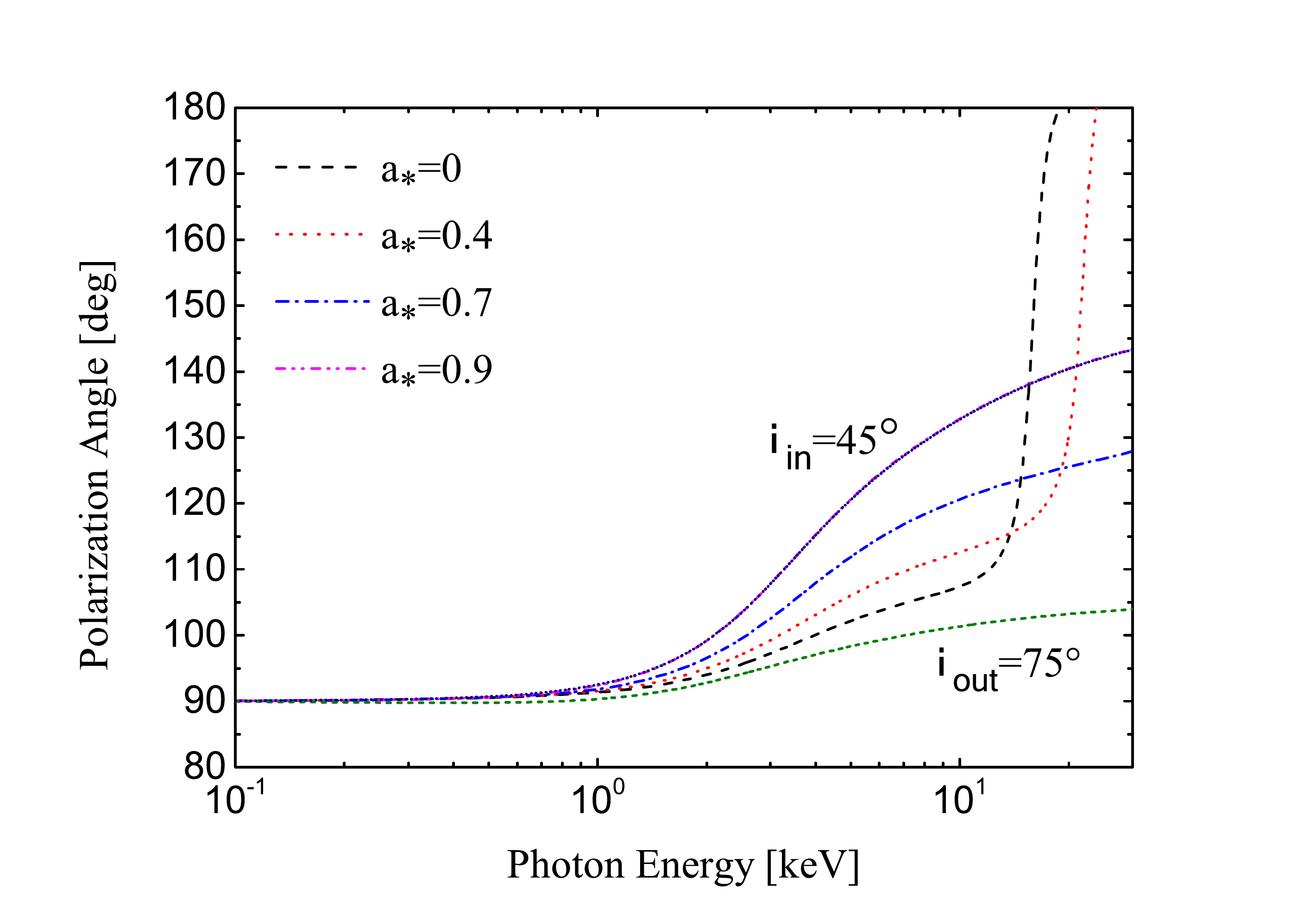}
\end{center}
\caption{As in the top panels in Fig.~\ref{fig3}, but now $R_* = 100$~$R_g$ is fixed and we consider different values of the spin parameters.}
\label{fighelp}
\end{figure*}

\section{Warped accretion disks \label{s-3}}

At this point, we want to compute the polarization properties of the thermal spectrum of a warped disk with warp radius $R_*$. Since there is no analytic model for warped disks in the literature, we use the following approach.

The inner part of the disk should be in the plane normal to the BH spin, as a consequence of the Bardeen-Petterson effect. We thus use a Novikov-Thorne model with inner radius at the ISCO, $R_{\rm ISCO}$, and outer radius the warp radius, $R_*$. The spacetime is described by the Kerr metric with spin parameter $a_* = a/M = J/M^2$, where $M$ and $J$ are, respectively, the mass and the spin angular momentum of the central BH. The inclination of the disk with respect to the line of sight of the distant observer is $i_{\rm in}$, where the subindex ``in'' refers to the fact we are considering the inner part of the disk.

The outer part of the disk is described by another Novikov-Thorne disk. Its inner edge is $R_*$ and the outer edge is some large radius $R_{\rm out}$ ($R_{\rm out} = 10^5$~$R_g$ in our simulations). We use the Schwarzschild metric to describe the outer disk. The reason for this is that there is no simple analytic description for a disk outside the equatorial plane in a Kerr metric. This simplification is justified since the outer disk is far from the compact object, $R_* \gg R_g$, and therefore the effects of the spin are negligible. We define the viewing angle of the outer disk as $i_{\rm out}$.

The inner and the outer disk are thus independent and the disk plane changes abruptly at the transition radius. We note that there is some uncertainty on the expected value of the possible warp radius, but a reasonable estimate is $R_* \sim 100$~$R_g$~\cite{r_w}.

If we neglect the radiation emitted from one of the disks, scattered by the other disk, and then detected by the observer, the system is completely specified by the viewing angles of the two disks with respect to the line of sight of the distant observer. In fact, there is an infinite number of configurations, because any axis of the disk can precess around the line of sight of the distant observer, changing the relative positions of the two disks. However, only the two viewing angles matter. The situation is different if we take into account that some radiation emitted from one of the disks can be scattered by the other disk, and in this case the polarimetric measurement at infinity does depend on the exact configuration of the system.

With this model for a warped disk, we have run our polarization code. As a first case, we assume the BH is not rotating. This is the simplest case and, in the absence of spin effects, cannot be expected to show strong effects since in the case of a non-rotating BH there is no Bardeen-Petterson effect. Figs.~\ref{fig1} and \ref{fig1w} show the spectrum of the observed polarization degree (left panels) and polarization angle (right panels) in the case of a disk with the inner part with an inclination angle $i_{\rm in} = 45^\circ$, while the outer part has $i_{\rm out} = 75^\circ$. The misalignment is thus $30^\circ$.

We include the effect of reflected radiation in the following manner. As described in Sec.~\ref{s-2}, the analysis is done by tracing the path of the photon in reverse. After the calculation of the photon trajectory from the distant observer to a point on the outer disk, we fire (isotropically with respect to the rest-frame of the accreting gas) photons from this point. We then include the contribution from those photons that subsequently land on the inner disk. The Stokes parameters $I$, $Q$, and $U$ are calculated from Tab.~XXV in~\cite{chandra}.

The results are plotted in Figs.~\ref{fig1} and~\ref{fig1w}. Fig.~\ref{fig1} includes only direct radiation, whereas Fig.~\ref{fig1w} includes both direct and reflected radiation. These plots show the polarization degree and angle for an unwarped Novikov-Thorne disk with $i = 45^\circ$ and $75^\circ$ together with those for a warped disk with warp radius $R_*/R_g = 30$, 100, 300, and 1000. Here and in what follows, as default values we assume a 10~$M_{\odot}$ BH with a luminosity of 10\% the Eddington limit.

In Figs.~\ref{fig2} and \ref{fig2w}, we show the case in which $i_{\rm in} = 75^\circ$ and $i_{\rm out} = 45^\circ$, without and with the reflected radiation, respectively. From Figs.~\ref{fig1} to~\ref{fig2w}, showing results in a Schwarzschild background, there is no appreciable difference between the calculations without and with the reflected radiation.

Roughly speaking, the high energy part of the spectrum of a warped disk looks like that of an unwarped disk with the same angle as the inner part of the warped disk, while the low energy part of the spectrum looks like that of an unwarped disk with the same angle as its outer disk. The radiation emitted by the inner disk and scattered by the outer disk introduces some distortion and complication. The presence of a warped disk leaves a signature in the spectrum of the polarization degree, while we are not able to say the same for the spectrum of the polarization angle.

As a second example, we have considered the same configurations, but now the spacetime of the inner disk is described by a Kerr metric with $a_* = 0.9$. The results are reported in Fig.~\ref{fig3} (hereafter all plots include the effect of reflected radiation). The effect of the spin is usually small and alters the spectrum at energies above a few keV, while the effect of a warped disk with $R_* > 30$~$R_g$ is at lower energies. In order to analyze these two features, we calculate the spectrum for a configuration with $R_* = 100$~$R_g$ and $a_* = 0$, 0.4, 0.7, and 0.9. Fig.~\ref{fighelp} shows these results. As the spin increases, the inner edge of the disk approaches the BH and there are more high energy photons. The impact of the reflected radiation is lower and moves to higher energies. In any case, the BH spin has no effects in the identification of the feature associated to the position of the warp radius. Moreover, as in the Schwarzschild case of Fig.~\ref{fig2w}, when $i_{\rm in} > i_{\rm out}$ the polarization degree can strongly decrease at lower energies, which cannot occur in the case of an unwarped disk. We note that the photon absorption due to the disk atmosphere decreases the polarization degree at lower energies, but it affects the spectrum at energies lower than about 0.1~keV~\cite{lixin}. Finally, these differences occur at a few KeV and therefore outside the region of interest to identify a warped disk.

\begin{figure*}[t]
\begin{center}
\includegraphics[type=pdf,ext=.pdf,read=.pdf,width=8.9cm]{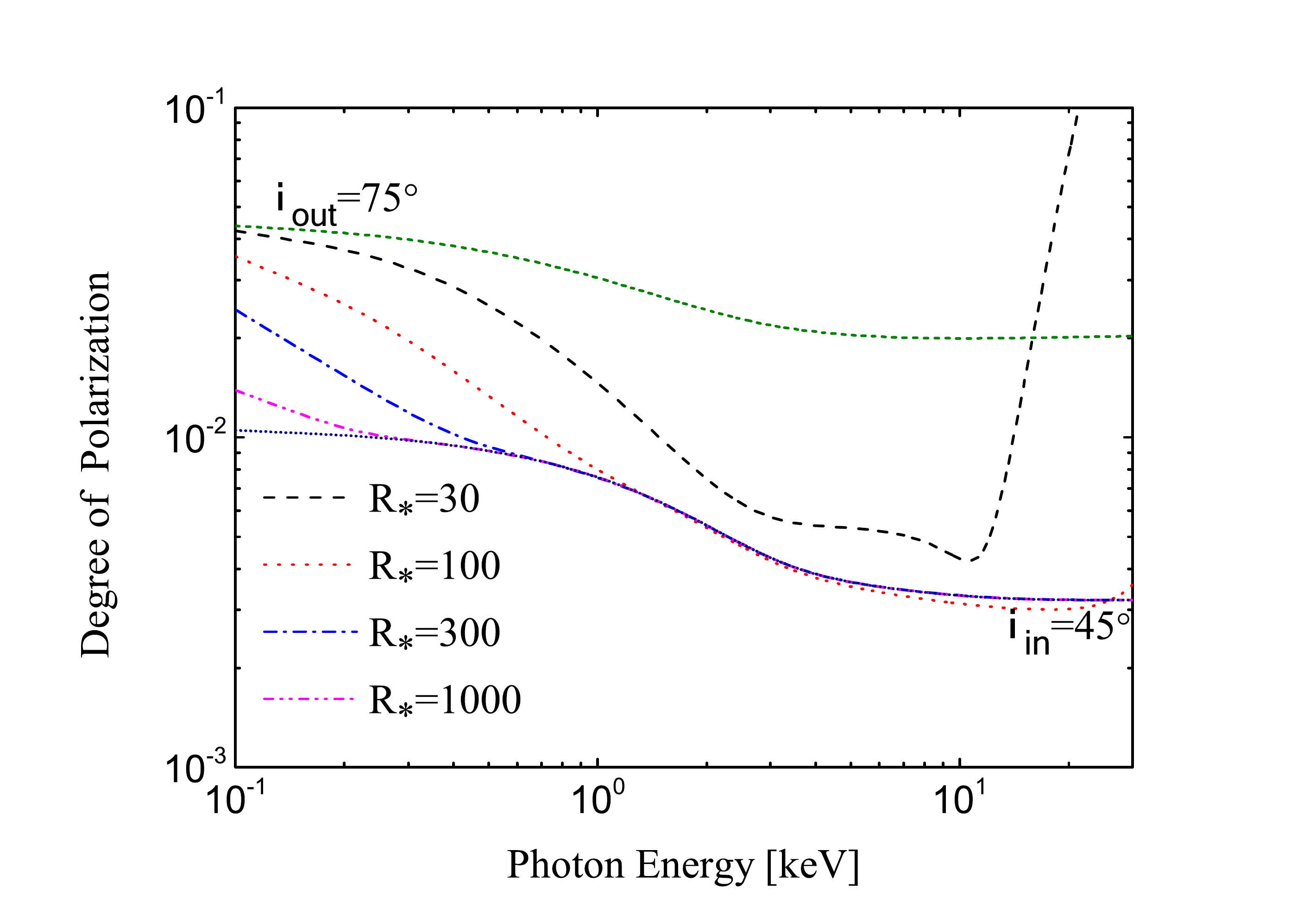}
\includegraphics[type=pdf,ext=.pdf,read=.pdf,width=8.9cm]{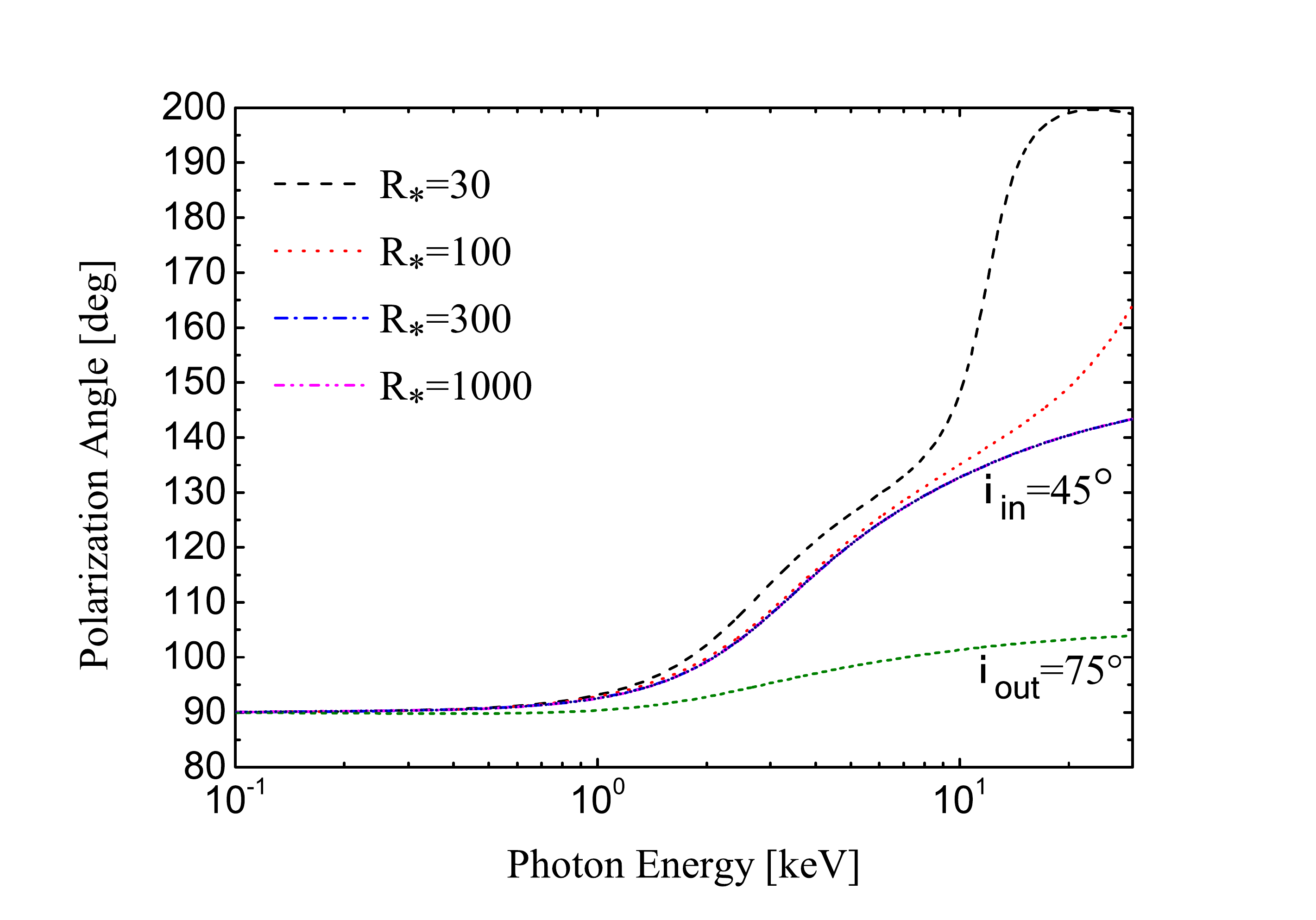}
\end{center}
\caption{As in the top panels in Fig.~\ref{fig3}, but after a rotation of $180^\circ$ of the axis of the inner disks about the line of sight of the distant observer.}
\label{figother}
\end{figure*}

As already mentioned, $i_{\rm in}$ and $i_{\rm out}$ do not completely specify the configuration of the system, because the axes of the two disks can rotate about the line of sight of the distant observer preserving the values of $i_{\rm in}$ and $i_{\rm out}$, but changing the relative position of the two disk. This, in turn, affects the corrections due to the radiation emitted by the inner disk and scattered by the outer disk. If we rotate one of the axes of the accretion disks, we get a different system configuration with a slightly different polarimetric spectrum. Fig.~\ref{figother} shows the counterpart of the top panels in Fig.~\ref{fig3} in which the axis of one of the two disk has been rotate by $180^\circ$, so the axes of the disk are on different semi-planes with respect to the line of sight of the distant observer. The rotation may have an effect on the measurement of the polarization degree and polarization angle, but the signature of a warped disk in the polarization degree is unaffected.

\begin{figure*}
\begin{center}
\includegraphics[type=pdf,ext=.pdf,read=.pdf,width=8.9cm]{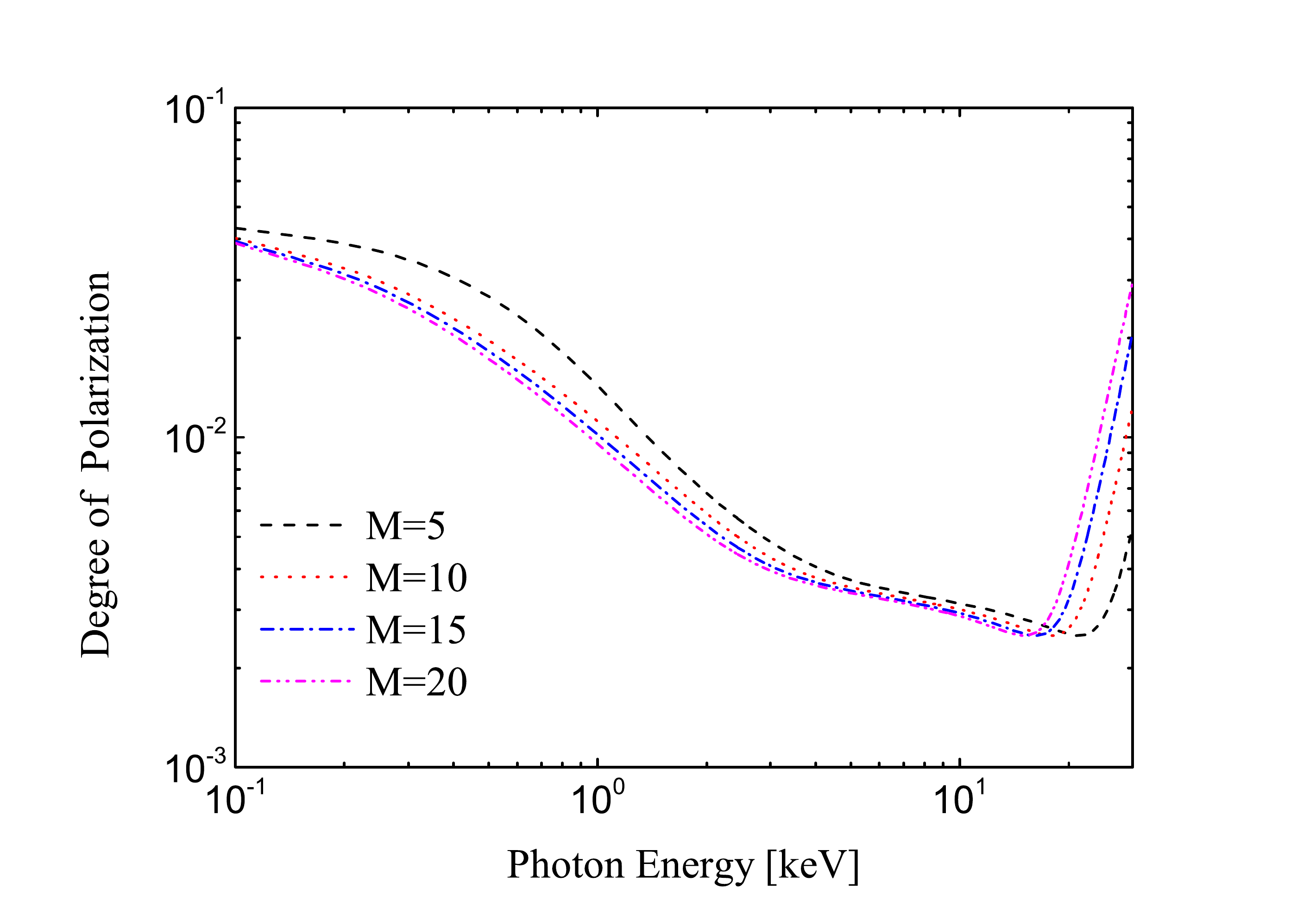}
\includegraphics[type=pdf,ext=.pdf,read=.pdf,width=8.9cm]{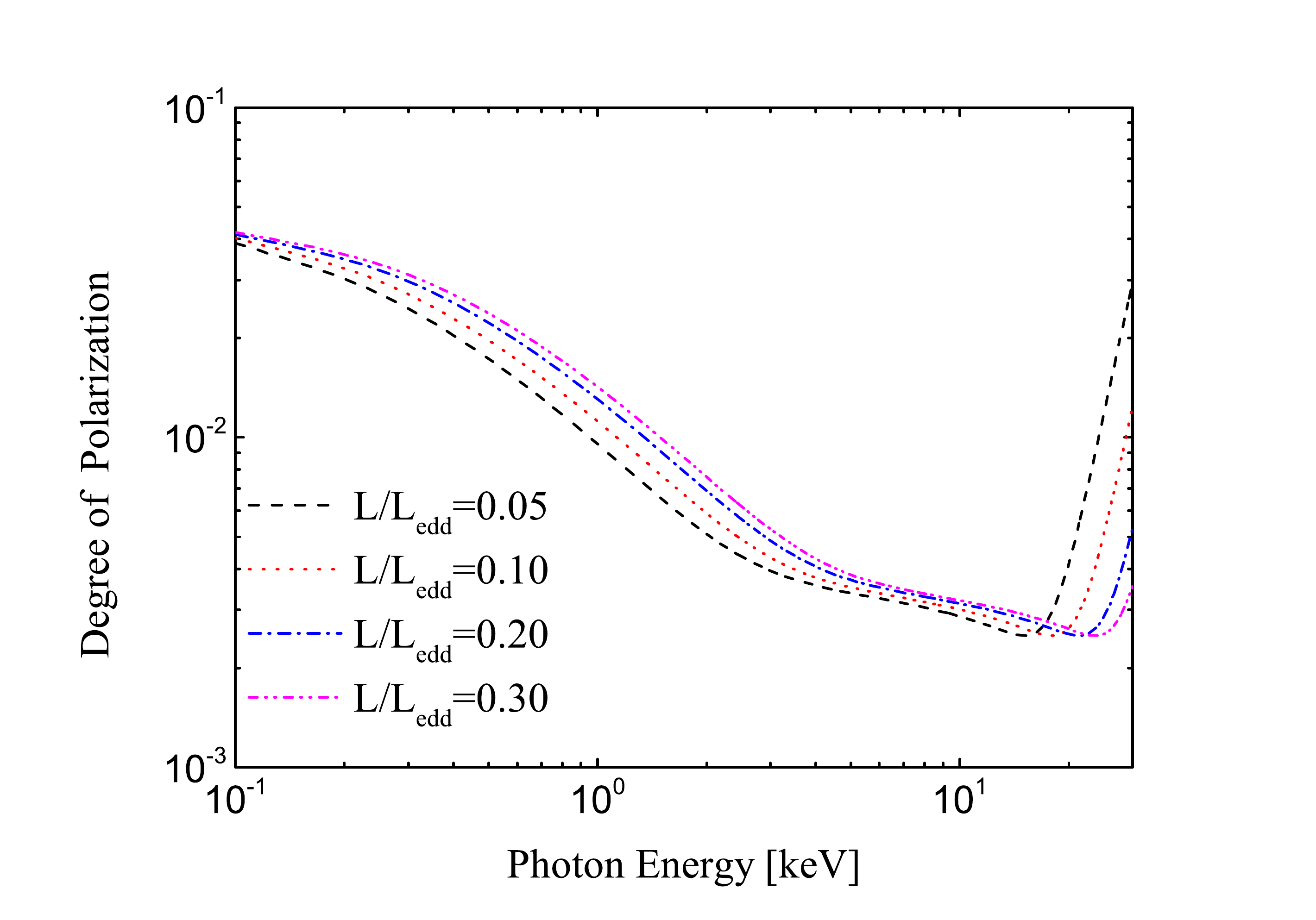}
\end{center}
\caption{Impact of the mass and of the accretion luminosity on the polarization degree for the thermal spectrum of an accretion disk around a Kerr BH with spin parameter $a_* = 0.9$. The inner part of the disk has an inclination angle $i_{\rm in} =45^\circ$ and the outer part has an inclination angle $i_{\rm out} =75^\circ$. The warp radius is at 50~$R_g$. Mass in Solar mass units, luminosity in terms of the Eddington luminosity.}
\label{fig4}
\vspace{0.3cm}
\begin{center}
\includegraphics[type=pdf,ext=.pdf,read=.pdf,width=8.9cm]{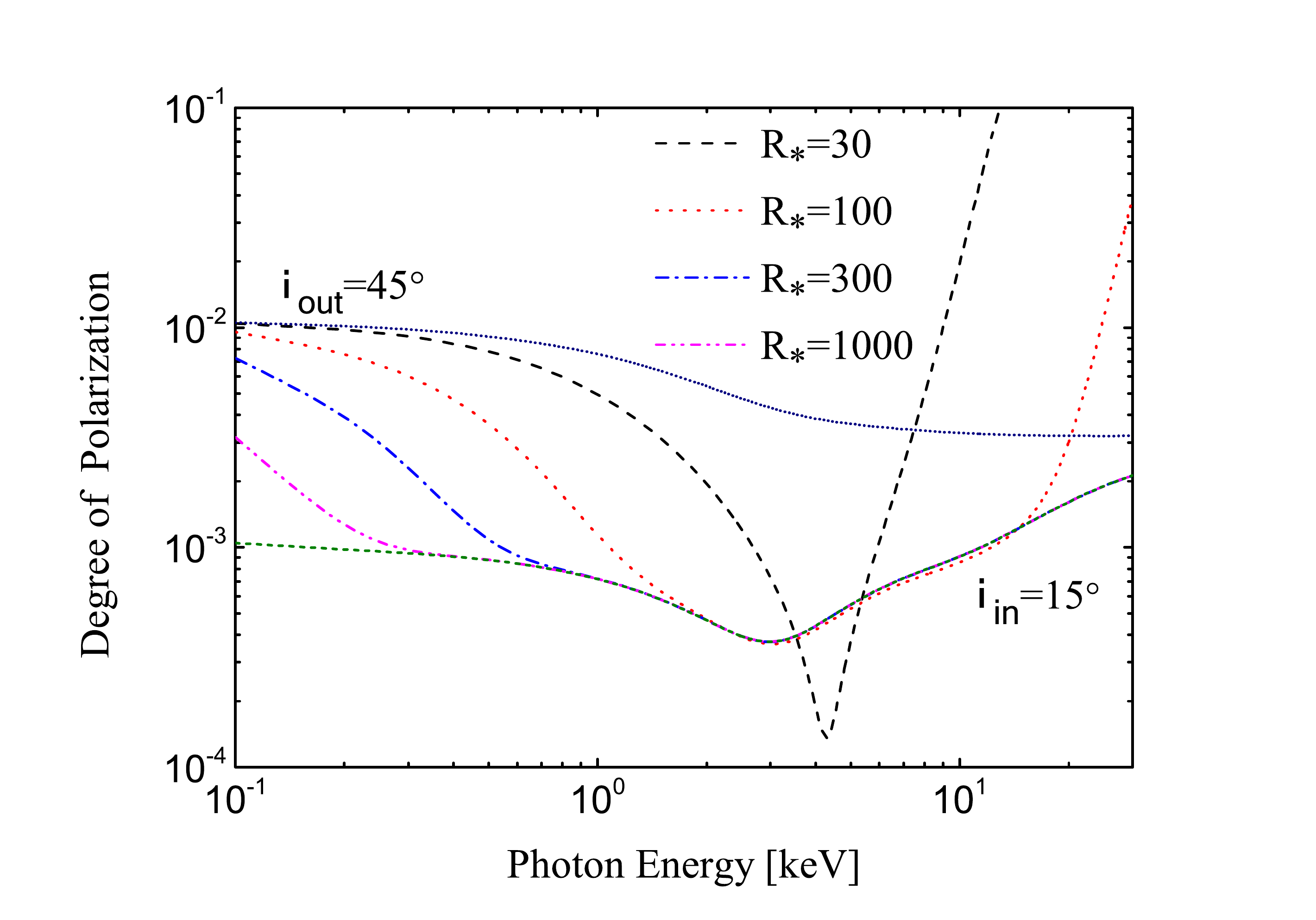}
\includegraphics[type=pdf,ext=.pdf,read=.pdf,width=8.9cm]{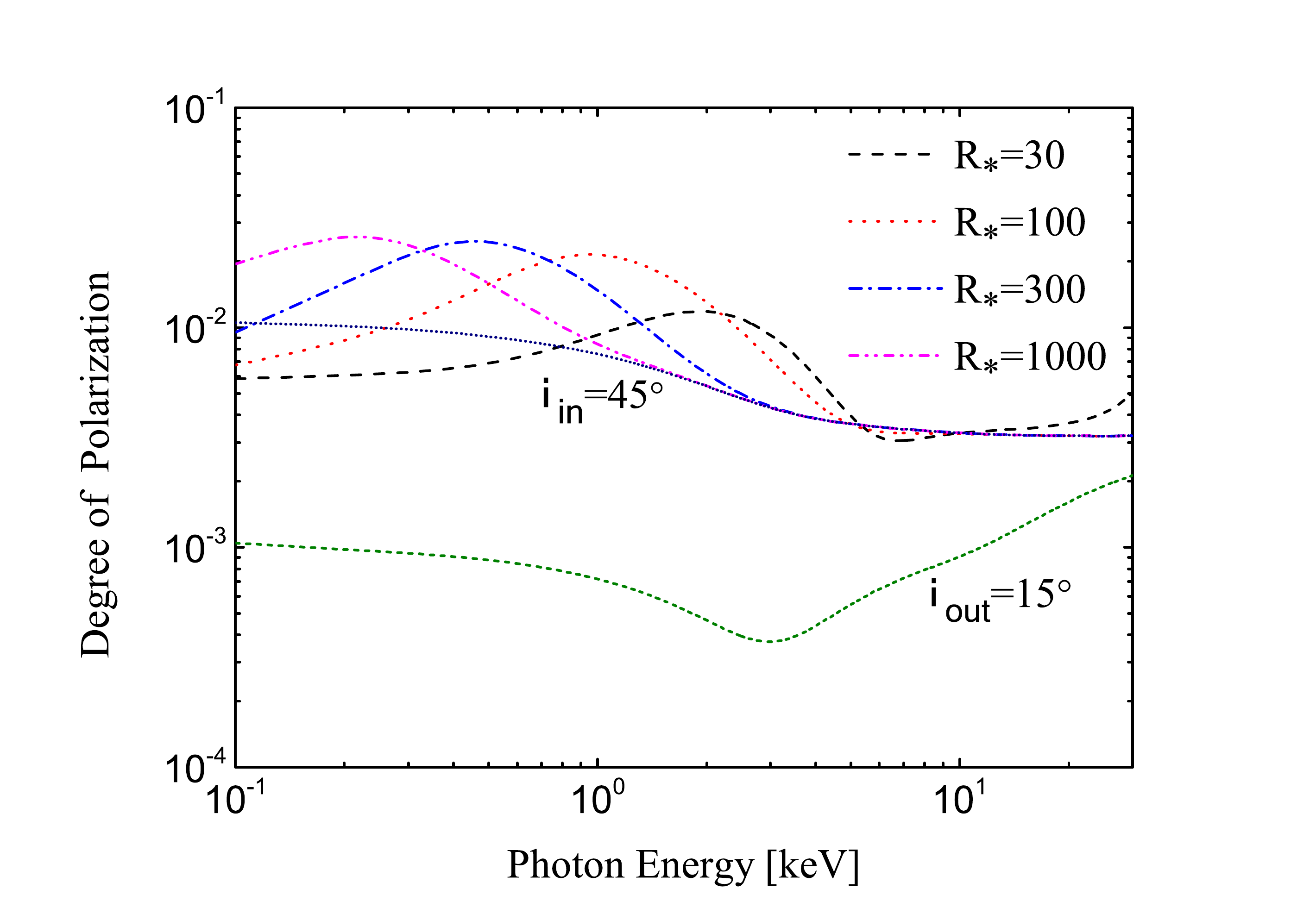}
\end{center}
\caption{Polarization degree for the thermal spectrum of an accretion disk around a Kerr BH with spin parameter $a_* = 0.9$. In the left panel, the inner part of the disk has an inclination angle $i_{\rm in} =15^\circ$ and the outer part has an inclination angle $i_{\rm out} =45^\circ$, while in the right panel we have the opposite case, namely the inner part of the disk has an inclination angle $i_{\rm in} =45^\circ$ and the outer part has $i_{\rm out} =15^\circ$. The radiation emitted from the inner disk and scattered by the outer disk is taken into account. See the text for more details.}
\label{fig5}
\end{figure*}

In Fig.~\ref{fig4}, we show the effect of the BH mass and accretion luminosity on the polarization degree. In these plots, the central BH has a spin $a_* = 0.9$, the inner disk has $i_{\rm in} = 45^\circ$, the outer part has $i_{\rm out} = 75^\circ$, and the warp radius is $R_* = 50$~$R_g$. Since a lower (higher) BH mass and a higher (lower) mass accretion rate have the effect to increase (decrease) the disk's temperature, the transition of the two regimes moves to higher (lower) energies. However, the effect is small.

Fig.~\ref{fig5} shows the spectrum of the polarization degree for a warped disk with other values of the inner and outer inclination angles. We do not show the spectrum of the polarization angle because, as evident from the previous figures, it has no unambiguous signature of the presence of a warped disk. Here the inclination angles are $i_{\rm in} = 15^\circ$ and $i_{\rm out} = 45^\circ$ in the left panel, and $i_{\rm in} = 45^\circ$ and $i_{\rm out} = 15^\circ$ in the right panel. In the former case, the feature associated to the warped disk is even more evident, because of the low polarization degree at high energies and therefore the polarization from the outer part of the disk starts dominating at higher energies. In the right panel, the polarization degree at low energies still receive a significant contribution from the inner disk and therefore the warped disk signature is somewhat different.

\section{Concluding remarks \label{s-4}}

X-ray polarization observations have the potentiality to explore the geometry of BH systems and they should be particularly suitable to infer the inclination angle of accretion disk around stellar-mass  BHs~\cite{lixin,schnittman}. Such observations are not possible with current instruments because they are much more challenging than similar observations in radio and optical bands. However, upcoming missions, e.g., eXTP~\cite{xtp}, XIPE~\cite{xipe}, IXPE~\cite{ixpe}, and PRAXYS~\cite{praxys} are promising tools for such observations in future.

The common picture of a BH in an X-ray binary is that the BH spin is perpendicular to the accretion disk and parallel to the orbital angular momentum of the stellar companion. Such an assumption plays a crucial role, for instance, in the spin measurement with the continuum-fitting method, in which the inclination angle of the disk must be an input parameter and it is usually inferred by optical measurements of the secondary star. Theoretical considerations suggest that both aligned and misaligned BH spins are possible. Current attempts to test the orientation of the BH spin with the orbital angular momentum of the binary rely on the measurement of the inclination of the jet, assuming that the latter is parallel to the BH spin. This assumption is not completely justified since we do not know the origin of the formation of jets and there is no commonly agreed perspective in literature~\cite{jets}. In presence of a misalignment, we should expect that the disk is warped at some radius $R_* \sim 100$~$R_g$. The outer part of the disk is necessarily in the orbital plane of the binary because it is formed by the material of the stellar companion, while the inner part of the disk should adjust in the BH equatorial plane due to the Bardeen-Petterson effect.

In this paper, we have computed the polarization of the thermal radiation of a warped accretion disk around a BH within a simple disk model. We find that if the disk is warped there is a clear observational signature in the spectrum of the polarization degree in the range 0.1-3~keV, assuming a BH with $M = 10$~$M_\odot$, with an accretion luminosity of 10\% the Eddington limit, and with a warp radius in the range 30-300~$R_g$. We observe that the spectrum at lower energies typically reflects the inclination angle of the outer disk, while at higher energies it is usually determined by the inclination angle of the inner disk and the reflected radiation. In the spectrum of the polarisation degree, we see this transition at 0.1-3~keV. In the case of a reasonable difference between the two inclination angles say, 30 degrees, this feature is usually quite prominent and easy to identify with a qualitative analysis. Furthermore,  we observe a small effect associated with the BH mass and luminosity on the transition between the two regimes. Since a lower (higher) BH mass and a higher (lower) mass accretion rate have the effect of increasing (decreasing) the disk's temperature, the transition between the two regimes moves to higher (lower) energies. The quantitative details depend, among other things, on the actual properties of the warped disk (inclination angles of the inner and outer part, value of the warp radius) and the capability of the polarimetric detector (sensitivity, energy band).

The first generation of X-ray polarimetric detectors will probably work in the energy range 1-10~keV, and our analysis shows that the warped disk signature can be observed in the case of warp radius below $100$~$R_g$ ($R_* < 100$~$R_g$). One important phenomenon that we have not included in the present analysis is the returning radiation, but since it is expected to have the strongest effect at high energies, we expect that the qualitative results obtained here will continue to remain valid in the presence of returning radiation. Thus, future detections of X-ray polarizations promise to be a useful tool 
for understanding accretion disks around BHs and the nature of these BHs. 


\begin{acknowledgments}
This work was supported by the NSFC (grants 11305038 and U1531117), the Shanghai Municipal Education Commission grant for Innovative Programs No.~14ZZ001, the Thousand Young Talents Program, and Fudan University. C.B. acknowledges also support from the Alexander von Humboldt Foundation.
\end{acknowledgments}


\end{document}